\documentclass[12pt]{article}
\usepackage{graphicx}
\begin{document}

\begin{center}

{\Large \bf Eugene Paul Wigner's Nobel Prize}

\vspace{5mm}
Y. S. Kim\\
Department of Physics, University of Maryland,  \\
College Park, Maryland 20742, U.S.A.\\
e-mail: yskim@umd.edu \\

\vspace{2mm}

\end{center}

\begin{abstract}

In 1963, Eugene Paul Wigner was awarded the Nobel Prize in Physics for
his contributions to the theory of the atomic nucleus and the
elementary particles, particularly through the discovery and application
of fundamental symmetry principles.   There are no disputes about this
statement.  On the other hand, there still is a question of why the statement
did not mention Wigner's 1939 paper on the Lorentz group, which was regarded
by Wigner and many others as his most important contribution in physics.
By many physicists, this paper was regarded as a mathematical exposition having
nothing to do with physics.  However, it has been more than one half century
since 1963, and it is of interest to see what progress has been made toward
understanding physical implications of this paper and its historical role in
physics.  Wigner in his 1963 paper defined the subgroups of the Lorentz group
whose transformations do not change the four-momentum of a given particle, and
he called them the little groups.  Thus, Wigner's little groups are for internal
space-time symmetries of particles in the Lorentz-covariant world.  Indeed, this
subgroup can explain the electron spin and spins of other massive particles.
However, for massless particles, there was a gap between his little group
and electromagnetic waves derivable Maxwell's equations.  This gap was not
completely removed  until 1990. The purpose of this report is to review the
stormy historical process in which this gap is cleared.  It is concluded that
Wigner's little groups indeed can be combined into one Lorentz-covariant formula
which can dictate the symmetry of the internal space-time time symmetries of
massive and massless particles in the Lorentz covariant world, just like
Einstein's energy-momentum relation applicable to both slow and massless
particles.

\end{abstract}

\vspace{10mm}

\newpage

\section{Introduction}\label{intro}

Let us start with Isaac Newton.  He formulated his  gravity law applicable to
two point particles.  It took him 20 years to extend his law for solid spheres
with non-zero radii.

\par
\begin{figure}[thb]
\centerline{\includegraphics[scale=3.2]{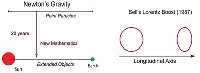} }
\caption{Newton's gravity law for point particles and extended objects.
It took him 20 years to formulate the same law for extended objects.
As for the classical picture of Lorentz contraction of the electron
orbit in the hydrogen atom, it is expected that the longitudinal component
becomes contracted while the transverse components are not affected.  In the
first edition of his book published in 1987, 60 years after 1927, John S.
Bell included this picture of the orbit viewed by a moving observer~\cite{bell04}.
While talking about quantum mechanics in his book, Bell overlooked the fact
that the electron orbit in the hydrogen atom had been replaced by a standing
wave in 1927.  The question then is how standing waves look to moving
observers.}\label{newton}
{}\hrulefill{}
\end{figure}

In 1905, Albert Einstein formulated his special relativity and  was
interested in how things look to moving observers.  He met Niels Bohr
occasionally.  Bohr was interested in the electron orbit of the hydrogen
atom.  Then, they could have talked about how the orbit looks to a moving
observer~\cite{bell04}, as illustrated in Fig.~\ref{newton}.   If they did,
we do not know anything about it.  Indeed, it is for us to settle this
Bohr-Einstein issue.

The purpose of the present paper is to discuss whether Wigner's 1939
paper on the Lorentz group provides the framework to address the internal
space-time symmetries of particles in the Lorentz-covariant world.  This
question is far more important than whether Wigner deserved a Nobel prize
for this paper alone.

For many years since 1963, many people  claimed that Wigner's 1939 paper
is worthless because he did not get the Nobel prize for it.  Let us
respond to this fatalistic view.  Einstein did not get the prize for his
formulation of special relativity in 1905.  Does this mean that Einstein's
special relativity worthless?  We shall return to this question in the
Appendix.

However, it is quite possible that Wigner started subject, but did not
finish it.  If so, how did this happen?
In his 1939 paper~\cite{wig39}, Wigner considered the subgroups of
the Lorentz group whose transformations leave the four-momentum of
a given particle invariant.  These subgroups are called Wigner's
little groups and dictate the internal space-time symmetries in
the Lorentz-covariant world.

He observed first that a massive particle at rest has three rotational
degree of freedom leading to the concept of spin.  Thus the little group
for this massive particle is like $O(3)$.  How about this massive
particle moving in the $z$ direction.  We could settle this issue
easily.

Wigner observed also that a massless particle cannot be brought to
its rest frame, but he showed that the little group for the massless
particle also has three degrees of freedom, and that this little
is locally isomorphic to the group $E(2)$ or the two-dimensional
Euclidean group.  This means that
generators of this little group share the same set of closed
commutation relations with that for two-dimensional Euclidean
group with one rotational and two translational degrees of freedom.

It is not difficult to associate the rotational degree of freedom
of $E(2)$ to the helicity of the massless particle.  However, what
is the  physics of the those two translational degrees of freedom?
Wigner did not provide the answer to this question in his 1939
paper~\cite{wig39}.  Indeed, this question has a stormy history,
and the issue was not completely settled until 1990~\cite{kiwi90jmp},
fifty one years after 1939, or 27 years after his Nobel prize in
1963.

\begin{figure}
\centerline{\includegraphics[scale=2.5]{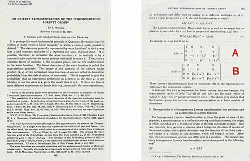}}
\caption{Wigner's 1939 paper in the Annals of Mathematics.  Its front page
is on page 149 of the journal.  On page 165, there are two strange matrices.
The second matrix (matrix B) is for the Lorentz boost along the $z$
direction.  However, the physics of the first matrix (matrix A) was not
completely understood until 1990, 51 years after 1939. }\label{annals}
{}\hrulefill{}
\end{figure}

\begin{table}
\begin{center}
\caption{One little group for both massive and massless particles.
Einstein's special relativity gives one relation for both.  Wigner's
little group unifies the internal space-time symmetries for massive and
massless particles which are locally isomorphic to $O(3)$ and $E(2)$
respectively. This table is from Ref.~\cite{hks86jmp}.}\label{gen11}
\vspace{1mm}
\begin{tabular}{cccc}
{}&{}&{}&{}\\ \hline \\[-4.8mm] \hline
{}&{}&{}&{}\\
{} & Massive, Slow \hspace*{6mm} & COVARIANCE \hspace*{6mm}&
Massless, Fast \\[2mm]\hline
{}&{}&{}&{}\\
Energy- & {}  & Einstein's & {} \\[-0.2mm]
Momentum & $E = p^{2}/2m$ & $ E = \sqrt{p^{2}+ m^{2}} $ &
$E = p$ \\[4mm]\hline
{}&{}&{}\\
Internal & $S_{3}$ & {}  &  $S_{3}$ \\[-0.2mm]
space-time &{} & Wigner's  & {} \\[-0.2mm]
symmetry & $S_{1}, S_{2}$ & Little Group &
Gauge Transformations\\
{}&{}&{}\\[2mm] \hline \\[-4.8mm] \hline
\end{tabular}
\end{center}
{}\hrulefill{}
\end{table}

For many years, the major complaint had been that his little groups
could not explain the Maxwell field.  Is it possible to construct
electromagnetic four-potential and the Maxwell tensor as representations
of representations of Wigner's little group for massless particles?

To answer this question, let us go to one of his matrices in his paper,
given in Fig.~\ref{annals}.   It is easy to see that matrix B is for a
Lorentz boost along the $z$ direction.   Matrix A leaves the
four-momentum of the massless particle invariant.  What else does this
matrix do?  In 1972~\cite{kuper76},
Kuperzstych showed that it performs a gauge transformation when applied to
the electromagnetic four-potential, but he did not see this as Wigner's
problem.  Indeed, this question was not completely answered until
1990~\cite{kiwi90jmp}.

In the present paper, we point out the complete understanding of this matrix
leads to result given in Table~\ref{gen11} contained the paper I published
with my younger colleagues in 1986~\cite{hks86jmp}.  As Einstein's
energy-momentum leads to its expressions both in the small-momentum
and large-momentum limits, Wigner's little groups explain the internal
space-time symmetries for the massive particle at rest as well as for
the massless particle, as summarized in Table~\ref{gen11}.

From Sec.~\ref{little} to Sec.~\ref{o3e2}, technical details are given.
The present author gave many lectures on this subject in the past.  In
this report, he explains the same subject at a public-lecture level by
avoiding group theoretical words as much as possible.   Since this paper
deals with a sensitive issue, it is appropriate to mention his background
and as well as his experience in dealing with those people who did not
agree with him.

In Sec.~\ref{little}, Wigner's little groups are spelled out in the
language of four-by-four matrices.
In Sec.~\ref{spinhalf}, the two-by-two representation if given for
spin-half particles.  The gauge transformation is defined for this
two-by-two representation.
In Sec.~\ref{fourvec}, it is shown that the gauge transformation defined
in the two-by-two representation leads to the gauge transformation applicable
to the four-potential for photons.
In Sec.~\ref{o3e2}, the $O(3)$-like little group for massive particles.

In the Appendix, it is noted that this paper deals with a very serious
historical issue.  The question then is whether the present author is
qualified to write about this issue.  Thus, it is appropriate to explain
how and why he had to study Wigner's paper.   A more interesting story
is how much resistance he had to face in making his results known to
Wigner and to the world.

\section{Wigner's little groups}\label{little}
If we use the four-vector convention $x^{\mu} = (x, y, z, t)$, the
generators of rotations around and boosts along the $z$ axis take the
form
\begin{equation}\label{eq001}
J_{3} = \pmatrix{0&-i&0&0\cr i&0&0&0\cr 0&0&0&0\cr 0&0&0&0} , \qquad
K_{3} = \pmatrix{0&0&0&0\cr 0&0&0&0 \cr 0&0&0&i \cr 0&0&i&0} ,
\end{equation}
respectively.  We can also write the four-by-four matrices for $J_{1}$
and $J_{2}$ for the rotations around the $x$ and $y$ directions, as
well as $K_{1}$ and $K_{2}$ for Lorentz boosts along the $x$ and $y$
directions respectively~\cite{knp86,bkn15}.  These six generators satisfy
the following set of commutation relations.
\begin{equation}\label{eq002}
\left[J_{i}, J_{j}\right] = i\epsilon_{ijk} J_{k}, \qquad
\left[J_{i}, K_{j}\right] = i\epsilon_{ijk} K_{k}, \qquad
\left[K_{i}, K_{j}\right] = -i\epsilon_{ijk} J_{k}.
\end{equation}
This closed set of commutation relations is called the Lie algebra of
the Lorentz group.   The three $J_{i}$ operators constitute a closed
subset of this Lie algebra.  Thus, the rotation group is a subgroup
of the Lorentz group.

In addition, Wigner in 1939 considered a subgroup generated
by~\cite{wig39}
\begin{equation}\label{eq005}
J_{3}, \qquad N_{1} = K_{1} - J_{2} ,\qquad N_{2} = K_{2} + J_{1} .
\end{equation}
These generators satisfy the closed set of commutation relations
\begin{equation}\label{eq200}
\left[N_{1}, N_{2}\right] = 0, \qquad
\left[J_{3}, N_{1}\right] = iN_{2}, \qquad
\left[J_{3}, N_{2}\right] = -iN_{1}.
\end{equation}

\begin{figure}
\centerline{\includegraphics[scale=1.8]{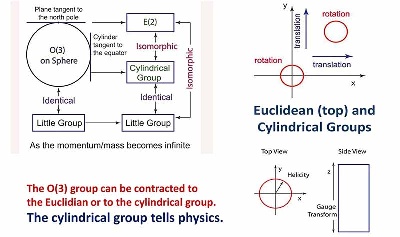}}
\caption{Transformations of the $E(2)$ group and the cylindrical
group.  They share the same Lie algebra, but only the cylindrical
group leads to a geometrical interpretation of the gauge
transformation.}\label{cylin}
{}\hrulefill{}
\end{figure}


As Wigner observed in 1939~\cite{wig39}, this set of commutation
relations is just like that for the generators of the two-dimensional
Euclidean group with one rotation and two translation generators,
as illustrated in Fig.~\ref{cylin}.  However, the question is what
aspect of the massless particle can be explained in terms of this
two-dimensional geometry.

Indeed, this question has a stormy history, and was not answered
until 1987.  In their paper of 1987~\cite{kiwi87jmp}, Kim and Wigner
considered the surface of a circular cylinder as shown in
Fig.~\ref{cylin}.  For this cylinder, rotations are possible around
the $z$ axis.  It is also possible to make translations along the
$z$ axis as shown in Fig.~\ref{cylin}.  We can write these generators as
\begin{equation}
L_{3} = \pmatrix{0 & -i & 0 \cr i & 0 & 0 \cr 0 & 0 & 0 }, \quad
Q_{1} = \pmatrix{0 & 0 & 0 \cr 0 & 0 & 0 \cr i & 0 & 0 }, \quad
Q_{2} = \pmatrix{0 & 0 & 0 \cr 0  & 0 & 0 \cr 0 & i & 0 } ,
\end{equation}
applicable to the three-dimensional space of $(x, y, z).$  They then
satisfy the closed set of commutation relations
\begin{equation}\label{eq300}
\left[Q_{1}, Q_{2}\right] = 0, \qquad
\left[L_{3}, Q_{1}\right] = iQ_{2}, \qquad
\left[L_{3}, Q_{2}\right] = -iQ_{1}.
\end{equation}
which becomes that of Eq.(\ref{eq200}) when $Q_{1}, Q_{2}, $ and
$L_{3}$ are replaced by $N_{1}, N_{2},$ and $J_{3}$ of Eq.(\ref{eq005})
respectively.  Indeed, this  cylindrical group is locally isomorphic
to Wigner's little group for massless particles.

Let us go back to the generators of Eq.(\ref{eq005}).  The role of
$J_{3}$ is well known.  It is generates rotations around the momentum and
corresponds to the helicity of the massless particle.
The $N_{1}$ and $N_{2}$ matrices take the form~\cite{knp86,bkn15}
\begin{equation}\label{eq6}
N_{1} = \pmatrix{0&0&-i&i\cr 0&0&0&0 \cr i&0&0&0 \cr i&0&0&0} , \qquad
N_{2} = \pmatrix{0&0&0&0 \cr 0&0&-i&i \cr 0&i&0&0 \cr 0&i&0&0} .
\end{equation}
The transformation matrix is
\begin{equation}\label{eq101}
D(u,v) = \exp{\left\{-i\left(uN_{1} + vN_{2}\right)\right\}}
 = \pmatrix{1 & 0 & -u & u \cr 0 & 1 & -v & v \cr
u & v & 1 - (u^{2}+ v^{2})/2 & (u^{2} + v^{2})/2 \cr
u & v & -(u^{2} + v^{2})/2 & 1  + (u^{2} + v^{2})/2} .
\end{equation}
In his 1939 paper~\cite{wig39}, Wigner observed that this matrix leaves
the four-momentum of the massless particle invariant as can be seen from
\begin{equation}\label{eq102}
\pmatrix{1 & 0 & -u & u \cr 0 & 1 & -v & v \cr
u & v & 1 - (u^{2}+ v^{2})/2 & (u^{2} + v^{2})/2 \cr
u & v & -(u^{2} + v^{2})/2 & 1  + (u^{2} + v^{2})/2}
\pmatrix{0 \cr 0 \cr p_{3} \cr p_{3}} = \pmatrix{0 \cr 0 \cr p_{3} \cr p_{3}} ,
\end{equation}
but he never attempted to apply this matrix to the photon four-potential.

\par
It is interesting to note that  in 1976 noted that this
form in applicable to the four-potential while making rotation and boosts
whose combined effects do not change the four four-momentum~\cite{kuper76}
of the photon.  In 1981, Han and Kim carried out the same
calculation within the framework of Wigner's little group~\cite{hk81ajp}.
Kuperzstych's conclusion was that the four-by-four matrix of Eq.(\ref{eq101})
performs a gauge transformation when applied to the photon four-potential,
and Han and Kim arrived at the same conclusion.  Let us see how this
happens.

\par

Let us next consider the  electromagnetic wave propagating
along the $z$ direction:
\begin{equation}
A^{\mu}(z,t) = (A_{1}, A_{2}, A_{3}, A_{0}) e^{i\omega (z - t)} ,
\end{equation}
and apply the $D(U,v)$ matrix to this electromagnetic four-vector:
\begin{equation}\label{eq110}
\pmatrix{1 & 0 & -u & u \cr 0 & 1 & -v & v \cr
u & v & 1 - (u^{2}+ v^{2})/2 & (u^{2} + v^{2})/2 \cr
u & v & -(u^{2} + v^{2})/2 & 1  + (u^{2} + v^{2})/2}
\pmatrix{A_{1} \cr A_{2} \cr A_{3} \cr A_{0}} ,
\end{equation}
which becomes
\begin{equation}\label{eq120}
\pmatrix{1 & 0 & 0 & 0 \cr 0 & 1 & 0 & 0 \cr u & v & 1  & 0 \cr u & v & 0 & 1}
\pmatrix{A_{1} \cr A_{2} \cr A_{3} \cr A_{0}}  -
\left(A_{3} - A_{0}\right)
\pmatrix{u \cr v \cr (u^{2}+ v^{2})/2 \cr (u^{2}+ v^{2})/2}.
\end{equation}
If the four-vector satisfies the Lorentz condition $A_{3} = A_{0}$,
this expression becomes
\begin{equation}\label{eq150}
\pmatrix{1 & 0 & 0 & 0 \cr 0 & 1 & 0 & 0 \cr u & v & 1 & 0 \cr
u & v & 0 & 1}\pmatrix{A_{1} \cr A_{2} \cr A_{3} \cr A_{0}}
=  \pmatrix{A_{1} \cr A_{2} \cr  A_{3}  \cr A_{0}}
+  \pmatrix{0 \cr 0 \cr  u A_{1} + v A_{3} \cr u A_{1} + v A_{3}   } .
\end{equation}
The net effect is an addition of the same quantity to the longitudinal and
time-like components while leaving the transverse components invariant.
Indeed, this is a gauge transformation.

\section{Spin-1/2 particles}\label{spinhalf}
Let us go back to the Lie algebra of the Lorentz group given in Eq.(\ref{eq002}).
It was noted that there are six four-by-four matrices satisfying nine
commutation relations.  It is possible to construct the same Lie algebra with
six two-by-two matrices~\cite{knp86,bkn15}.  They are
\begin{equation}
J_{i} = \frac{1}{2} \sigma_{i}, \quad\mbox{and}\quad
    K_{i} = \frac{i}{2} \sigma_{i} ,
\end{equation}
where $\sigma_{i}$ are the Pauli spin matrices.  While $J_{i}$ are Hermitian,
$K_{i}$ are not.  They are anti-Hermitian.  Since the Lie algebra of Eq.(\ref{eq002})
is Hermitian invariant, we can construct the same Lie algebra with
\begin{equation}
J_{i} = \frac{1}{2} \sigma_{i}, \quad\mbox{and}\quad
    \dot{K}_{i} = -\frac{i}{2} \sigma_{i} .
\end{equation}
This is the reason why the four-by-four Dirac matrices can explain both
the spin-1/2 particle and the anti-particle.
\par

\begin{figure}
\centerline{\includegraphics[scale=1.0]{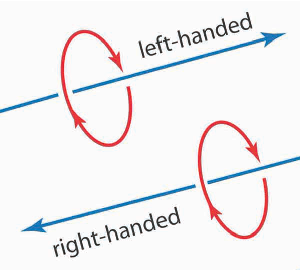}}
\caption{Polarization of massless neutrinos.  Massless neutrinos are
left-handed, while anti-neutrinos are right-handed.
This is a consequence of gauge invariance.}\label{neupol}
{}\hrulefill{}
\end{figure}
Thus the most general form of the transformation matrix takes the form
\begin{equation} \label{eq705}
T = \exp{\left( -\frac{i}{2}\sum_{i}\theta_{i}\sigma_{i}
+ \frac{1}{2}\sum_{i}\eta_{i}\sigma_{i}\right)} ,
\end{equation}
and this transformation matrix is applicable to the spinors
\begin{equation}\label{eq701}
\chi_{+} = \pmatrix{1 \cr 0}, \quad\mbox{and}\quad \chi_{-} = \pmatrix{0 \cr 1},
\end{equation}
In addition, we have to consider the transformation matrices
\begin{equation}
\dot{T} = \exp{\left( -\frac{i}{2}\sum_{i}\theta_{i}\sigma_{i}
  - \frac{1}{2}\sum_{i}\eta_{i}\sigma_{i}\right)} ,
\end{equation}
applicable to
\begin{equation}\label{eq702}
\dot{\chi}_{+} = \pmatrix{1 \cr 0}, \quad\mbox{and}\quad
   \dot{\chi}_{-} = \pmatrix{0 \cr 1}.
\end{equation}

\par
With this understanding, let us go back to the Lie algebra of Eq.(\ref{eq002}).
Here again the rotation generators satisfy the closed set of commutation
relations:
\begin{equation}
\left[J_{i}, J_{j}\right] = i\epsilon_{ijk} J_{k},  \qquad
\left[\dot{J}_{i}, \dot{J}_{j}\right] = i\epsilon_{ijk} \dot{J}_{k} .
\end{equation}
These operators generate the rotation-like $SU(2)$ group, whose physical
interpretation is well known, namely the electron and positron spins.

\par

Here also we can consider the $E(2)$-like subgroup generated by
\begin{equation}
J_{3}, \qquad N_{1} = K_{1} - J_{2}, \qquad N_{2} = K_{2} + J_{1} .
\end{equation}
The $N_{1}$ and $N_{2}$ matrices take the form
\begin{equation}
N_{1} = \pmatrix{0 & i \cr 0 & 0}, \qquad N_{2} = \pmatrix{0 & 1 \cr 0 & 0} .
\end{equation}

\par
On the other hand, in the ``dotted'' representation,
\begin{equation}
\dot{N}_{1} = \pmatrix{0&0 \cr -i & 0} , \qquad
\dot{N}_{2} = \pmatrix{0&0\cr 1&0}.
\end{equation}
There are therefore two different $D$ matrices:
\begin{equation}\label{d201}
D(u,v) = exp{\left\{-\left(iuN_{1} + ivN_{2}\right)\right\}}
 = \pmatrix{1 & u - iv \cr 0 & 1} ,
\end{equation}
and
\begin{equation} \label{d202}
\dot{D}(u,v) = exp{\left\{-\left(iu\dot{N}_{1} + iv\dot{N}_{2}\right)\right\}}
  = \pmatrix{1 & 0 \cr u + iv & 1} .
\end{equation}
These are the gauge transformation matrices applicable to  massless spin-1/2
particles~\cite{bkn15,hks82}.

Here are talking about the Dirac equation for with four-component spinors.

The spinors $\chi_{+}$ and $\dot{\chi}_{-}$ are gauge-invariant since
\begin{equation}
         D(u,v)\chi_{+} =\chi_{+}, \quad\mbox{and}\quad
         \dot{D}(u,v)\dot{\chi}_{-} = \dot{\chi}_{-} .
\end{equation}
As for $\chi_{-}$ and $\dot{\chi}{+}$,
\begin{eqnarray}\label{eq706}
&{}& D(u,v) \chi_{-} = \chi_{-} + (u - iv)\chi_{+} , \nonumber \\[2mm]
&{}& \dot{D}(u,v) \dot{\chi}_{+} =
\dot{\chi}_{+} + (u + iv)\dot{\chi}_{-} .
\end{eqnarray}
They are not invariant under the $D$ transformations, and they are not
gauge-invariant.  Thus, we can conclude that the polarization of
massless neutrinos is a consequence of gauge invariance, as illustrated in
Fig.~\ref{neupol}.

\section{Four-vectors from the spinors}\label{fourvec}

We are familiar with the way in which the spin-1 vector is constructed
from the spinors in non-relativistic world. We are now interested in
constructing four-vectors from these spinors.  First of all,
with four of the spinors given above, we can start with
the products.
\begin{equation}
 \chi_{i}\chi_{j}, \quad \chi_{i} \dot{\chi}_{j},  \quad
 \dot{\chi}_{i}\chi_{j},   \quad  \dot{\chi}_{i}\dot{\chi}_{i} .
\end{equation}
resulting in spin-0 scalars and four-vectors and four-by-four tensors for
the spin-1 states~\cite{bkn15}.  The four-vector can be constructed from
the combinations $\chi_{i} \dot{\chi}_{j}$ and $\dot{\chi}_{i}\chi_{j}.$

\par
Among them, let us consider the combinations, let us consider the
four resulting from $\dot{\chi}_{i}\chi_{j}$.  Among them,
As far as the rotation subgroup is concerned, $\dot{\chi}_{+}\chi_{+}$, and
$\dot{\chi}_{-}\chi_{-}$ are like $-(x + iy)$ and $(x - iy)$ respectively, and
and invariant under Lorentz boosts along the $z$ direction.  In addition,
we should consider
\begin{equation}
  \frac{1}{2}\left(\dot{\chi}_{-}\chi_{+} +
                                   \dot{\chi}_{+}\chi_{-}\right) ,
 \quad\mbox{and}\quad
  \frac{1}{2}\left(\dot{\chi}_{-}\chi_{+}  -
                                   \dot{\chi}_{+}\chi_{-}\right),
\end{equation}
which are invariant under rotations around the $z$ axis.  When the system
boosted along the $z$ direction, these combinations are transformed like
$z$ and $t$ directions respectively.

\par
With these aspects in mind, let us consider the matrix
\begin{equation}
 M = \pmatrix{ \dot{\chi}_{-}\chi_{+} & \dot{\chi}_{-}\chi_{-} \cr
    - \dot{\chi}_{+}\chi_{+}  & - \dot{\chi}_{+}\chi_{-} } ,
\end{equation}
and write the transformation matrix $T$ of Eq.(\ref{eq705})as
\begin{equation}
 T = \pmatrix{\alpha & \beta \cr \gamma & \delta} , \quad{\mbox{with}} \quad
 \det{(T)} = 1 .
\end{equation}
If four matrix elements are complex numbers, there are
eight independent parameters.  However, the condition $ \det{(T)} = 1 $
reduces this number to six.  The Lorentz group starts with six degrees of freedom.

It is then possible to write the four-vector $(x, y, z, t)$ as
\begin{equation}\label{eq707}
   X = \pmatrix{t + z & x - iy \cr x + iy & t - z} ,
\end{equation}
with its Lorentz-transformation property
\begin{equation} \label{eq717}
X' = T~X~T^{\dag} ,
\end{equation}
The four-momentum can also be written as
\begin{equation}
   P = \pmatrix{p_{0} + p_{3} & p_{1} - ip_{2} \cr
        p_{1}+ ip_{2} & p_{0}  - p_{3}} ,
\end{equation}
with the the transformation property same as that for $X$ given in
Eq.(\ref{eq717}).

\par
With this understanding, we can write the photon four-potential as
\begin{equation}
 A = \pmatrix{A_{0} + A_{3} & A_{1} - iA_{2}  \cr
 A_{1} + iA_{2} & A_{0} - A_{3} }
\end{equation}
Let us go back the two-by-two matrices $D(u,v)$ and $\dot{D}(u,v)$ given in
Eqs.(\ref{d201}) and (\ref{d202}).  We said there that they perform gauge
transformations on massless neutrinos. It is indeed gratifying to note that
they also lead to the gauge transformation applicable to the photon
four-potential.

\begin{equation}
   D(u,v) A D^{\dag}(u, v) =  \pmatrix{ 1 & u - iv \cr 0 & 1}
\pmatrix{A_{0} + A_{3} & A_{1} - iA_{2}  \cr
 A_{1} + iA_{2} & A_{0} - A_{3} }  \pmatrix{ 1 & 0  \cr u + iv  & 1}  .
\end{equation}
This results in
\begin{equation}
 \pmatrix{A_{0}+ A_{3}  + 2(u A_{1} + v A_{2}) &
         A_{1} - iA_{2}  \cr
         A_{1} + iA_{2}+  &  A_{0} - A_{3}}
   + (A_{0} - A_{3})\pmatrix{  u^2 + v^2  & u - iv \cr u + iv &  1}.
\end{equation}
If we apply the Lorentz condition $A_{0} = A_{3} $, this matrix becomes
\begin{equation}
\pmatrix{2A_z + 2(u A_{1} + v A_{2}) & A_{1} - iA_{2} \cr
         A_{1} + iA_{2} & 0}.
\end{equation}
This result is the same as the gauge transformation in the four-by-four
representation given in Eq.(\ref{eq150}).

\section{Massless particle as a limiting case of a massive
  particle}\label{o3e2}

In his 1939 paper~\cite{wig39}, Wigner discussed his little groups for
massive and massless particles as two distinct mathematical devices.
Indeed, In{\"o}n{\"u} and Wigner in 1953 initiated of the unification of
these little groups by observing
considering a flat plane tangent to a sphere, while the plane and sphere
correspond to the $E(2)$ and $O(3)$ symmetries respectively~\cite{inonu53}.
This unification was completed in 1990~\cite{kiwi90jmp}.  The issue is
whether the $E(2)$-like little group can be obtained as a zero-mass limit of
the $O(3)$-like little group for massive particles.  Another version of this
limiting process is given in Sec.~\ref{o3e2} of the present report.

\par
As for the internal space-time symmetry of particles, let us go beck to Bohr and
Einstein.  Bohr was interested in the electron orbit of the hydrogen atom while
Einstein was worrying about how things look to moving observers. They
met occasionally before and after 1927 to discuss physics.  Did they talk about
how the stationary hydrogen atom would look to a moving observer?  It they did,
we do not know about it.

This problem is not unlike the case of Newton'a law of gravity.  Newton worked out
the inverse square law for two point particles.  It took him 20 years to work
out the same law for extended objects such and the sun and earth, as illustrated
in Fig.~\ref{newton}.

In 1905, Einstein formulated his special relativity for point particles.  It is
for us to settle the issue of how the electron orbit of the hydrogen atom looks to
moving observers.  Indeed,  the circle and ellipse as given in Fig.~\ref{newton}
have been used to illustrate this this relativistic effect.  However, these figures
do not take into account the fact that the electron orbit had been replaced by a
standing wave.  Indeed, we should learn how to Lorentz-boost standing waves.

\par
Yes, we know how to construct standing waves for the hydrogen atom.  Do we know
how to Lorentz-boost this atom?  The answer is No.  However, we can replace
it with the proton without changing quantum mechanics.  Both the hydrogen atom
and the proton are quantum bound states, but the proton can be accelerated.
While the Coulomb force is applicable to the hydrogen, the harmonic oscillator
potential is used as the simplest binding force for the quark model~\cite{fkr71}.
We can switch the Coulomb wave functions with oscillator wave functions without
changing quantum mechanics.
This problem is illustrated in Fig.~\ref{quapar}.  Then it is possible to
construct the oscillator wave functions as a representation of Wigner's little
group~\cite{knp86,bkn15,kno79jmp}.
In this two-by-two representation, the Lorentz boost along the positive
direction is
\begin{equation}
     B(\eta) = \pmatrix{e^{\eta/2} & 0 \cr 0 & e^{-\eta/2}},
\end{equation}
the rotation around the $y$ axis is
\begin{equation}
R(\theta) = \pmatrix{\cos(\theta/2) & -\sin(\theta/2) \cr
                  \sin(\theta/2) & \cos(\theta/2)} .
\end{equation}
Then, the boosted rotation matrix is
\begin{equation}\label{eq607}
B(\eta) R(\theta) B(-\eta) =
         \pmatrix{\cos(\theta/2) & - e^{\eta}\sin(\theta/2) \cr
                  e^{-\eta}\sin(\theta/2) & \cos(\theta/2)} .
\end{equation}

If $\eta$ becomes very large, and this matrix is to remain finite,
$\theta$ has to become very small, and this expression becomes~\cite{kmn16}
\begin{equation}
 \pmatrix{1 - r^2 e^{-2\eta}/2 &  r \cr
                 - r e^{-2\eta} & 1 - r^2 e^{-2\eta}/2} .
\end{equation}
with
\begin{equation}\label{gamma}
 r = - \frac{1}{2}\theta e^{\eta} .
 \end{equation}
This expression becomes
\begin{equation}
D(r) = \pmatrix{1 &  r \cr 0 & 1} .
\end{equation}
In this two-by-two representation, the rotation around the $z$ axis is
\begin{equation}
Z(\phi) = \pmatrix{e^{-i\phi/2} & 0 \cr 0 & e^{i\phi/2}}, \qquad
\end{equation}
respectively.  Thus
\begin{equation}
  D(u,v) = Z(\phi) D(r) Z^{-1}(\phi) ,
\end{equation}
which becomes
\begin{equation}\label{eq711}
       D(u, v) = \pmatrix{1 & u - iv \cr 0 & 1} ,
\end{equation}
with
\begin{equation}\label{eq602}
u = r\cos\phi, \quad\mbox{and}\quad  v = r\sin\phi,
\end{equation}

\par
Here, we have studied how the little group for the $O(3)$-like little  group
the massive particle becomes the $E(2)$-like little group for the massless
particle in the infinite-$\eta$ limit.   What does this limit mean
physically?     The parameter $\eta$ can be derived from the speed of
of the particle.  We know $\tanh(\eta) = v/c$, where $v$ is the speed of
the particle.  Then
\begin{equation}
\tanh\eta = \frac{p}{\sqrt{m^2 + p^2}},
\end{equation}
where $m$ and $p$ are the mass and the momentum of the particle respectively.
If $m$ is much smaller than $/p$,
\begin{equation}
          e^{\eta} = \frac{\sqrt{2} p}{m} ,
\end{equation}
which becomes large when $m$ becomes very small.  Thus, the limit of large
$\eta$ means the zero-mass limit.


Let us carry our the same limiting process for the  four-by-four
representation.  From the generators of the Lorentz group, it is possible
to construct the four-by-four matrices for rotations around the $y$ axis
and Lorentz boosts along the $z$ axis as~\cite{bkn15}
\begin{equation}
 R(\theta) = \exp{\left(-i\theta J_{2}\right)}, \quad\mbox{and}\qquad
 B(\eta) = \exp{\left(-i\eta K_{3}\right)},
\end{equation}
respectively.  The Lorentz-boosted rotation matrix is $B(\eta) R(\theta) B(-\eta)$
which can be written as
\begin{equation}\label{eq600}
\pmatrix{\cos\theta  & 0 & (\sin\theta)\cosh\eta & -(\sin\theta)\sinh\eta \cr
         0 & 1 & 0  & 0 \cr
       -(\sin\theta)\cosh\eta & 0 & \cos\theta - (1 - \cos\theta)\sinh^2\eta &
        (1 - \cos\theta)(\cosh\eta) \sinh\eta \cr
       -(\sin\theta)\cosh\eta & 0 & -(1 - \cos\theta)(\cosh\eta)\sinh\eta &
        \cos\theta + (1 - \cos\theta)\cosh^{2}\eta} .
\end{equation}
While $\tanh\eta = v/c$, this boosted rotation matrix becomes a transformation
matrix for a massless particle when $\eta$ becomes infinite. On the other hand,
it the matrix is to be finite in this limit, the angle $\theta$ has to become
small.  If we let $r = - \frac{1}{2}\theta e^{\eta}$ as given in Eq.(\ref{gamma}),
this four-by-four matrix becomes
\begin{equation}\label{eq601}
\pmatrix{1  & 0 & -r &  r \cr
         0 & 1 & 0  & 0 \cr
       r & 0 & 1 - r^2/2 &  r^2/2  \cr
       r & 0 & -r^2/2 & 1 + r^2/2} .
\end{equation}
This is the Lorentz-boosted rotation matrix around the $y$ axis.  However,
we can rotate this $y$ axis around the $z$ axis by $\phi$.  Then
the matrix becomes
\begin{equation}\label{eq603}
\pmatrix{1  & 0 & -r \cos\phi & r \cos\phi \cr
         0 & 1 & -r \sin\phi  & r\sin\phi \cr
       r \cos\phi & r \sin\phi & 1 - r^2/2 &  r^2/2  \cr
       r \cos\alpha & r \sin\phi & -r^2/2 & 1 + r^2/2} .
\end{equation}
This matrix becomes $D(u,v)$ of Eq.(\ref{eq101}), if replace $r \cos\phi$
and $r \sin\phi$ with $u$ and $v$ respectively, as given in Eq.(\ref{eq602}).

\begin{appendix}

\section{Author's Qualifications}
In this report, I am dealing with a very serious issue in physics.  The question
is whether I am qualified to talk about Wigner's 1939 paper.  The reader of this
article is not likely to be the first one to raise this issue.

\begin{figure}
\centerline{\includegraphics[scale=2.5]{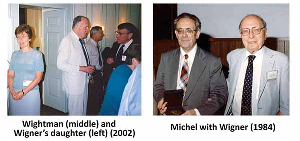}}
\caption{Arthur Wightman and Louis Michel.  Wightman's photo was taken at
a luncheon meeting held at Princeton University to celebrate Wigner's
100-th birth day in 2002.  Michel's photo was taken during the 13th
International Colloquium on Group Theoretical Methods in Physics held at
the University of Maryland in 1984.  Both photos were taken by the author
of this paper.}\label{micwi}
{}\hrulefill{}
\end{figure}

Louis Michel and Arthur Wightman were among the most respected physicists on the
Lorentz group, and their photos are in Fig.~\ref{micwi}.  In 1961, while I was a
graduate student at Princeton, I was in Wightman's class.  I learned
from him the ABC of the Lorentz group.  Wightman gave the same set of lectures
in France, and he published an article in French~\cite{wight62}.

In 1962, Louis Michel gave a series of very comprehensive lectures at a conference
held in Istanbul. Indeed, I learned from his lecture note~\cite{michel62} how the
inhomogeneous Lorentz group is different from the homogenous Lorentz group.

Both Michel and Wightman became upset when I was meeting Wigner during the
period from 1985 to 1990.  Wightman sent me a letter telling me my papers
on Wigner's 1939 paper are wrong.  In particular, he said the table given in
Table~\ref{gen11} is wrong.  I assume he told the same story to Wigner
because his office and Wigner's office were in the same building on the
campus of Princeton University.

Louis Michel became very angry when I had to tell him I could carry out my
Wigner program without his permission. He told me he did not like what I say
in Table~\ref{gen11}.  He even wrote a letter to John S. Toll telling him to
reduce my position at the University of Maryland in 1987. Toll was the
chancellor of the state-wide University of Maryland system at that time.

\begin{figure}
\centerline{\includegraphics[scale=2.0]{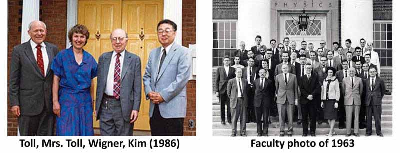}}
\caption{Toll, Mrs. Toll, Wigner, and Kim at the Chancellor's Mansion
of UMD (1986).  The physics faculty photo of UMD (1963).  Kim is the
youngest man standing in the middle of the second row.}\label{twk}
{}\hrulefill{}
\end{figure}

He was John A. Wheeler's student at Princeton University and came to the
University of Maryland in 1953 to build the present form of the physic
department.  In 1962, he hired me as an assistant professor one year after my
PhD degree at Princeton.  Toll became very happy whenever Wigner came to
Maryland at my invitation, as indicated in Fig.~\ref{twk}.

In spite of those hostile reactions from Michel and Wightman,  Wigner
liked Table~\ref{gen11}, and continued listening to me.  John S. Toll
continued supporting my position at the University of Maryland.  In spite
of what I said above I still like Michel and Wightman.  They were great
teachers to me.

\begin{figure}
\centerline{\includegraphics[scale=2.5]{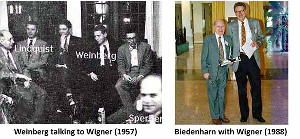}}
\caption{Steven Weinberg and Lawrence Biedenharn. They had their own
positive views toward Wigner's 1939 paper. In this photo, Weinberg is
talking to Wigner in 1957, when he was a graduate student at Princeton.
Biedenharn is standing with Wigner in 1988 during the first Wigner
Symposium held at the University of Maryland.}\label{biedenwein}
{}\hrulefill{}
\end{figure}


Stephen Adler and Gerald Feinberg were also very influential physicists
during the period from 1960 to 1990.  I knew them well.  In 1981, when I
submitted a paper with my younger colleagues on the Wigner issue to the
Physical Review Letters, Feinberg wrote a referee report saying Wigner's
1939 paper is a useless mathematical exposition having nothing to do with
physics.  He was so proud of what he was saying that he revealed his name
in his report, while the referees are anonymous.   Thus, he deserves to
be mentioned in the present report.

Since Feinberg did not give other reasons in his report, we resubmitted
the paper asking the editor to examine its scientific contents.  At that
time, Adler was in the editorial position to make the final decision.
Adler said he agreed with Feinberg without making any comments of his own.
In other words, Adler was also saying that Wigner's 1939 paper is worthless.
In effect, both Adler and Feinberg were telling us not to waste my time
because Wigner did not get the Nobel prize for this paper.

Steven Weinberg was different.  In 1964, he
published a series of papers on the spin states that can be constructed
from Wigner's little groups~\cite{wein64a,wein64b,wein64c}.  Indeed,
he realized that Wigner's little groups are for the internal space-time
symmetries.  As for massless particles, Weinberg realized the matrix A
of Fig.~\ref{annals} was troublesome, and constructed his ``state vectors''
which are independent of this matrix~\cite{wein64b,wein64c}.   Does
Weinberg's result bring Wigner's paper closer to the Maxwell theory?

In the Maxwell formalism, it is possible to construct gauge-independent
states, namely electromagnetic fields. It is also possible to construct
the electromagnetic four-potential which depends on gauge transformations.
Thus, it is not difficult to guess Weinbeg's state vectors are for the
electromagnetic field, while matrix A of Fig.~\ref{annals} is for gauge
transformations applicable to the four-potential~\cite{bkn15}.

\begin{figure}
\centerline{\includegraphics[scale=5.0]{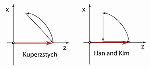}}
\caption{Lorentz transformations which leave the four-momentum
invariant.  However, these transformations do not lead to identity matrices.
When applied to the electromagnetic four-potential, they perform gauge
transformations.  Furthermore, since two-successive Lorentz boosts lead
to one boost preceded by a rotation (called Wigner rotation) in the
kinematics of Han and Kim, their transformation is the same as that of
Kuperzstych.}\label{loops}
{}\hrulefill{}
\end{figure}

With this point in mind, I published in 1981 a paper with Han saying
that  Matrix A  performs a gauge transformation~\cite{hk81ajp}.  We
considered a momentum-preserving transformation by considering one
rotation followed by two boosts, as shown in Fig.~\ref{loops}.  We
submitted this paper to the American Journal of Physics instead of the
Physical Review, because we felt that we were not the first ones to
observe this.  Indeed, in 1972, Kuperzstych got essentially the same
result, as indicated also in Fig.~\ref{loops}.  It is remarkable that
he got this result without making reference to Wigner's 1939 paper.
He concluded his paper saying that the concept of spin could be generated
from his momentum-preserving transformation.  It is remarkable that he
derived this result without relying on the concept of the little groups
spelled out in Wigner's 1939 paper~\cite{wig39}.

In 1953,  In{\"o}n{\"u} and Wigner published a paper on group
contractions~\cite{inonu53}.  We can study the contraction of the $O(3)$
group to $E(2)$ by considering a sphere for $O(3)$ and a two-dimensional
plane for $E(2)$.  We can then
consider this plane tangent to the sphere at the north-pole.  If the radius
of the sphere becomes very large, the spherical surface at the north-pole
becomes flat enough to accommodate the $E(2)$ symmetry.  We can constrauct
a flat football field on the surface of the earth.  Thus, the $E(2)$-like
little group for massless particles can be obtained from the $O(3)$-like
little group for massive particles.  Then, what is the physics of the
large-radius limit?

In his 1939 paper~\cite{wig39}, Wigner considered the little group of the
massive particle at rest.  What is then the little group for a particle
moving along the $z$ direction.  The answer is very simple.  It is a
Lorentz-boosted rotation matrix.  What happens when the momentum becomes
infinite?

In order to address this question, we can start from Einstein's
$E = \sqrt{m^2 + p^2}$.  We all know the form of this relation in the
limit of small $p/m$.  We also know the form for the large-$p/m$ limit.
With this point in mind, with Han and Son, I published a paper telling
that the the rotational degrees of freedom around $x$ and $y$ directions
become one gauge degree of freedom while the rotation around the $z$
axis remains as the helicity degree of freedom, as $p/m$ becomes
infinite~\cite{hks83pl}.  After several stages of refinements, we
published Table~\ref{gen11} in the Journal of Mathematical Physics in
1986~\cite{hks86jmp}.

Wigner liked this table.  This is precisely the reason why I was able to
publish a number of papers with him.  However, Wigner pointed out to me
that the geometry of the two-dimensional plane cannot explain the gauge
transformation, as indicated in Fig.~\ref{cylin}.
We thus worked hard to figure out the solution to this problem.   For a
given sphere, we can consider also a cylinder tangential to the equatorial
belt, as well as a plane tangential to the north pole, as illustrated in
Fig.~\ref{cylin}.  We published this result in the Journal of Mathematical
Physics in 1987~\cite{kiwi87jmp}, and another paper in 1990~\cite{kiwi90jmp}.
Lawrence Biedenharn was the chief editor of the Journal. He was very happy
to publish these papers and congratulated me on reactivating Wigner's 1939
paper.

I am very happy to include in this report Biedenharn's photo with Wigner
in Fig.~\ref{biedenwein}, which I took with my Canon AE camera in 1988.
Included in the same figure is a photo of Weinberg talking to Wigner
while he was a graduate student at Princeton in 1957.  Dieter Brill
contributed this photo.  Weinberg was Sam Treiman's first PhD student
and got his degree in 1957.  Since I went to Princeton in 1958, I did not
have overlapping years with Weinberg, but I had to read his thesis to
copy his style of writing.  I still like his English style.

Of course, I am proud of working with Wigner during his late years.  On
the other and, could I do this job without my own background?   I had
to fix up Wigner's work in order to put my own physics program on a
solid theoretical ground.  When I was graduate student, and for several
years after my PhD degree, I lived in the world where the origin of
physics is believed to be in the complex plane of the S-matrix, and
bound states should be described by poles in the complex plane.

\begin{figure}
\centerline{\includegraphics[scale=2.7]{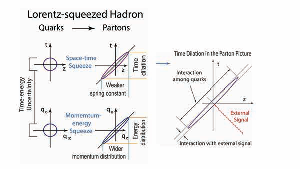}}
\caption{Gell-Mann's Quark model and Feynman's parton model as one
Lorentz-covariant entity.  The circle-ellipse diagram is from
Ref.~\cite{knp86}, and also from  Refs.~\cite{kim89}. This
Lorentz-squeeze is known as the space-time entanglement in the current
literature~\cite{bkn16}.}\label{quapar}
{}\hrulefill{}
\end{figure}

In 1965, when I pointed out those poles do not necessarily lead to
localized wave functions~\cite{ys04arx}, I had to face a stiff
resistance from the influential members of the American physics
community.  I choose not to mention their names.  They told me wave
functions have nothing to do with physics.  This is a totally
destructive attitude.  However, I took their reactions constructively.
We do not know how to Lorentz-boost bound-state wave functions,
while the S-matrix is a product of the Lorentz-covariant field theory.

\begin{table}[thb]
\begin{center}
\caption{One little group for both massive and massless particles.
In this table, we have added the last row to Table~\ref{gen11} telling
Gell-Mann's quark model and Feynman's parton model are two different
manifestations of one Lorentz-covariant entity.  This table is from
Ref.~\cite{knp86} and also from Ref.~\cite{kim89}.}\label{gen22}
\vspace{1mm}
\begin{tabular}{cccc}
{}&{}&{}&{}\\ \hline \\[-4.8mm] \hline
{}&{}&{}&{}\\
{} & Massive, Slow \hspace*{6mm} & COVARIANCE \hspace*{6mm}&
Massless, Fast \\[2mm]\hline
{}&{}&{}&{}\\
Energy- & {}  & Einstein's & {} \\[-0.2mm]
Momentum & $E = p^{2}/2m$ & $ E = \sqrt{p^{2}+ m^{2}} $ &
$E = p$ \\[4mm]\hline
{}&{}&{}\\
Internal & $S_{3}$ & {}  &  $S_{3}$ \\[-0.2mm]
space-time &{} & Wigner's  & {} \\[-0.2mm]
symmetry & $S_{1}, S_{2}$ & Little Group &
Gauge Transformations\\
{}&{}&{}\\[2mm] \hline
{}&{}&{}\\
Moving  &Gell-Mann's & Covariant & Feynman's  \\[-0.2mm]
H-atom & Quark Model & Bound State  & Parton Model \\
{}&{}&{}\\[2mm] \hline \\[-4.8mm] \hline

\end{tabular}
\end{center}
{}\hrulefill{}
\end{table}

Thus, my problem was to find at least one wave function that can be
Lorentz-boosted.  I then realized that Dirac in 1945~\cite{dir45} and
Yukawa in 1953~\cite{yuka53} constructed a Gaussian form that can be
Lorentz-boosted.

\par
In April of 1970, at the spring meeting of the American physical Society,
Feynman gave a talk where he repeatedly mentioned wave functions.  He was
talking about hadrons as bound states of quarks.  My colleagues were saying
Feynman was absolutely crazy, but he was a savior to me.  Let us face the
difficulty of boosting wave functions.  In 1971~\cite{fkr71}, with his
students, Feynman published a paper based on his 1970 talk~\cite{fkr71}.
There, they start with a Lorentz-invariant differential equation for
particles bound together by harmonic-oscillator potential.  However, they
produce solutions containing the Gaussian form
\begin{equation}
\exp{\left\{-\frac{1}{2}\left(x^2 + y^2 + z^2 - t^2 \right)\right\} } .
\end{equation}
This form is invariant under Lorentz transformations, but it is not possible
to physical interpretations to $t$ variable.

On the other hand, Feynman's differential equation also produces the
solutions containing thee Gaussian form
\begin{equation}\label{cov11}
\exp{\left\{-\frac{1}{2}\left(x^2 + y^2 + z^2 + t^2 \right)\right\} } .
\end{equation}
In this case, the wave function is normalizable in all the variables.  The
form is not invariant.  This means the wave function appears differently to
moving observers.  Figure~\ref{quapar} illustrates how differently the wave
function look differently.

With Marilyn Noz, I used the Gaussian form of Eq.(\ref{cov11}) to show
that Gell-Mann's quark model and Feynman's parton model are two different
limits of one Lorentz-covariant entity, and submitted our result to Physical
Review Letters.  The referee was very competent and honest.  He/she said
he/she really learned what the parton model is all about and the result is
important.  However, he would ``enthusiastically'' recommend publication
in the Comments and Addenda section of the Phys. Rev. D, instead of PRL.
We accepted his/her recommendation and published paper as a
comment~\cite{kn77}.

However, what aspect of the fundamental symmetry does  this quark-parton
reflect?  In order to answer this question, I had to study Wigner's 1939
paper, and show that the Lorentz-covariant oscillator wave functions are
representations of Wigner's $O(3)$-like little group~\cite{knp86,kno79jmp}.
My continued efforts led to a PRL paper of 1989~\cite{kim89}.  In that
paper, I expanded Table~\ref{gen11} to Table~\ref{gen22}.  This paper
also contains the major portion of Fig.~\ref{quapar}.  The elliptic
squeeze described in this fiigure is called the space-time entanglement
in the current literature~\cite{bkn16}.

Let me summarize what I said above.  Many people told me I am totally
isolated from the rest of the physics world while working on the problem
nobody else worries about. I disagree.  I have been in touch with all
essential physicists in this field, including Eugene Paul Wigner.   In
other words, I am qualified to write this report.

\end{appendix}

\end{document}